# Spreadsheet Components For All

*Jocelyn Paine*
www.j-paine.org/ *and* www.spreadsheet-parts.org/
popx@j-paine.org

**ABSTRACT**

*We have prototyped a "spreadsheet component repository" Web site, from which users can copy "components" into their own Excel or Google spreadsheets. Components are collections of cells containing formulae: in real life, they would do useful calculations that many practitioners find hard to program, and would be rigorously tested and documented. Crucially, the user can tell the repository which cells in their spreadsheet to use for a component's inputs and outputs. The repository will then reshape the component to fit. A single component can therefore be used in many different sizes and shapes of spreadsheet. We hope to set up a spreadsheet equivalent of the high-quality numerical subroutine libraries that revolutionised scientific computing, but where instead of subroutines, the library contains such components.*

**NOTE**

You can try a demonstration of the repository for Google Spreadsheets via the Web form at [Paine 2008b], www.j-paine.org/cgi-bin/google/filter_form.py. It is explained at [Paine 2008a], www.j-paine.org/dobbs/spreadlets.html.

## 1. INTRODUCTION

In everyday usage, a "component" is a part, usually manufactured. But the word often carries another implication. To an electronics engineer, a component is something such as a resistor, transistor or integrated circuit that you select from a catalogue. The catalogue documents the component's specification precisely enough that you are in no doubt about how to use it. If, having bought it, you find it fails to meet the spec, you can demand a replacement: in other words, the component comes with a guarantee. Should you need more components of a similar kind, you can buy them as readily as you bought the first: you're entitled to presume that if a supplier stocks a 5.6KΩ resistor, you'll have no trouble obtaining other standard values such as 4.7KΩ and 6.8KΩ. Finally, a component is manufactured via processes you probably can't invoke for yourself, usually involving extreme temperatures, exotically toxic metalloids, and abstruse quantum-physical calculations.

This, *mutatis mutandis*, is a spreadsheet component. Something that is to become part of your spreadsheet — i.e. a group of cells; that performs a calculation you neither know nor care how to program for yourself; that is comprehensively and comprehensibly documented; that has been tested so rigorously that the supplier feels safe in offering a no-quibble support agreement; and that admits all reasonable variations. Meaning that if its job is (say) to remove duplicates from an input range, you're entitled to presume that if it can accept rows as inputs, it can also accept columns as inputs, and probably any rectangle of cells on any sheet.

In the rest of this paper, I describe a prototype Web site for delivering such components to Google Spreadsheets [Google b] and Excel: a "spreadsheet parts repository". For Excel, it sends components as files representing groups of cells that can, with a suitable

VBA (Visual Basic for Applications) program, be copied into your Excel spreadsheet. For Google Spreadsheets, you identify your spreadsheet by filling in your Google Docs details and the spreadsheet name in a form. The repository will then copy the cell groups directly into that spreadsheet. You can try this via the form at [Paine 2008b].

Crucial is that although the repository sends components as groups of cells, it does not store them as such groups. Instead, it stores them as "templates", implemented in the Excelsior spreadsheet-generator language I describe in Sections 2.2 and 2.3. From these, it can generate many differently-shaped cell groups. This means that you can tell the repository which cells in your spreadsheet to use as a component's inputs and outputs. It will then apply this information to the appropriate template, in effect reshaping it to fit.

As far as I know, the Web-based repository I describe in this paper is the first such system. There are many other programs and utilities for generating Excel files, but none appear to have been used to deliver components in the way I propose. Probably the most important task remaining is to discover what components are most likely to benefit Excel users; and then, of course, to implement them. Then, I hope, the repository will do for spreadsheeting what high-quality numerical subroutine libraries such as the Numerical Algorithms Group [NAG] did for scientific computing. I also hope it will be possible to find permanent funding so that the repository will be free for all users.

### 1.1 Content of this paper

The next section gives background on the nature of components. Section 3 describes how the prototype Web site was implemented. Section 4 is a brief evaluation. Section 5 proposes future lines of study and implementation. Section 6 is a brief conclusion. Section 7 lists references and links. The Appendices contain technical material that may not interest all readers: examples of Excelsior for specifying and documenting components. Appendix 1 lists the Excelsior source of the component described in Section 2. Appendix 2 is a longer example showing how an unusual spreadsheet, one that generates science-fiction plots, can be specified and documented. The point is to show Excelsior-style documentation at work on a task that almost all readers will find unfamiliar. If the documentation makes it clear to you, the appendix will have helped to show that Excelsior is a viable documentation tool. Appendix 3 explains just enough of the Excelsior language for you to make sense of Appendices 1 and 2.

## 2. BACKGROUND

In this section, I explain what I mean by "spreadsheet component" (Section 2.1); how they would be used (Section 2.2); and how the repository stores them (Section 2.3).

### 2.1 What do I mean by "spreadsheet component"?

In the Introduction, I talked in very general terms about components. It's now time to examine the idea in more detail, and I shall describe what a typical component might look like, why it's the kind of thing worth making available, and why users might find it hard to program for themselves. As my example, I'll use a component for filtering elements out of tables, which I needed recently for creating dropdown menus.

Last year, I worked on a spreadsheet for modelling the finances of urban regeneration projects. These projects entailed building houses and flats, and users had to be able to input lists of dwelling types. Later on, the spreadsheet had to generate dropdown menus from these lists. Now, Excel allows a dropdown menu to take its options from a cell range. So, for instance, you can put in cell C7 of a sheet, a dropdown that takes its options from cells A1:A3 of the sheet. If these happen to contain the strings "House", "Flat" and "Bungalow", then the dropdown will display "House", "Flat" and "Bungalow" as options. We tried doing this with our spreadsheet, by associating the category names table with the dropdown. The problem was that because of how the tables were organised, some cells in the table were blank. This meant some of the options on the menu also became blank, which looked ugly and confused users. So we needed a way to remove these blanks, copying the remaining non-blank cells into a new table from which the dropdown would take its options.

The spreadsheet below illustrates almost this, except that to show the idea generalised to other patterns, it actually removes cells that don't start with X. Should you want to try it yourself, it is linked from [Paine 2007c]. This also links to the original of the component documentation displayed in Appendix 1.

**This is an example of the filter part that removes all strings that do not match a specified pattern. In this example, the pattern is X***

| elements to search | the index | matching elements |
|---|---|---|
| Not X | 2 | X |
| X | 5 | X2 |
| Not X | 10 | X4 |
| Not X | 11 | X5 |
| X2 | -1 | |
| Not X | -1 | |
| Not X | -1 | |
| Not X | -1 | |
| | -1 | |
| X4 | -1 | |
| X5 | -1 | |
| Not X | -1 | |
| Not X | -1 | |

The first two rows are text. The rest of column A is a is a table from which we want to copy all cells starting with X to column C, closing up gaps. In Excel, you can find such cells by a formula such as match( "X*", A3:A15, 0 ), the star being a wildcard. This is why the text in the first row talks about X*. Column B is a working table. The first cell of the table, B3, is the offset of the first cell in column A starting with an X, 2. (I.e. cell A4.)

The second cell of the table, B4, is the offset of the first cell starting with an X, 5. (I.e. cell A7.) The third cell of the table, B5, is the offset of the third cell starting with an X, 10. (I.e. cell A12.) And so on. The -1's indicate that there are no more such cells.

Now, it is possible to program this in Excel, without having to drop into VBA — which for various reasons, we didn't want to do. However, it is not easy. In the next paragraph, I'll explain a little why, so you have one example of *why* some components are hard to do in Excel.

Technically speaking, the natural way to program this is to use a technique called recursion, as so:

> Dear Excel,
>   this is how to remove blank cells from location i of table In onwards, putting the non-blank ones into table Out at location o onwards:
> **1.** If the cell at location i of In is blank, remove blank cells from location i+1 of table In onwards, putting the non-blank ones into table Out at location o onwards.
> **2.** If the cell at location i of In is not blank, copy it to location o of table Out. Then remove blank cells from location i+1 of table In onwards, putting the non-blank ones into table Out at location o+1 onwards.
> **3.** If there are no more cells at or after location i of In, stop.

This is a terribly familiar style of programming if you are a computer scientist: recursion is then something you'll have been taught in your first term, if not first week. However, if you're an Excel developer (unless perhaps you learnt programming in some other language first), it probably isn't. Recursion is normally used when defining functions; but Excel doesn't let you define functions (except in VBA). Even if it did, it's not good at data structures, and probably wouldn't have a notation for talking about tables and table segments in the way needed to translate the above rules.

**2.2 How are spreadsheet components used?**

I've explained why something might be worth regarding as a component. How should it be used? If you inspect the formulae in the Web copy of the spreadsheet above, linked from [Paine 2007c], you will see that there is nothing special about where the three tables are. We could write a similar spreadsheet with the "elements to search" table in column E instead of A, or column AJ, or even arranged along a row. We could put the "matching elements" table in column X, or row 19 of this sheet, or row 6 of any other sheet. Whatever the tables' relative positions and orientations, the formulae follow the same scheme.

Similarly, they follow the same scheme no matter how large the "elements to search" table.

What we need, therefore, is a way to encode this scheme in a way independent of the tables' sizes, orientations, and relative positions; and then to enable a user to take this encoding. And we need a way for a user to tell this encoding where in *their* spreadsheet the "elements to search" table (i.e. the inputs) is, and where the "matching elements" table (i.e. the outputs) is. Also to tell it whether the tables run horizontally or vertically, and how big they are. We can then substitute these values into the scheme, and obtain formulae appropriate to *their* spreadsheet. All we then need is to make these formulae available as an XLS file, or something else the user can copy into their spreadsheet.

This is what the Excelsior spreadsheet generator does. I described it in earlier papers [Paine 2005; Paine, Tek, Williamson 2006; Paine 2007a]; and explain briefly in Appendix 3 in case you want to read the code examples in Appendices 1 and 2. Here, therefore, I shall just explain the basic idea.

**2.3 How are spreadsheet components implemented? Excelsior in a nutshell**

Suppose we have the equation:

```
t[i] = u[i] * 2
```

Suppose also that t and u are both groups of 5 cells. Let t be cells A1:A5, and u be cells C1:G1. Assume the square brackets signify "array indexing" or "array subscripting". (If you've programmed in any conventional languages, you'll be familiar with this.) The idea is that the index i denotes an offset from the first cell in the table being indexed. Thus, t[1] is cell A1, t[2] is cell A2, t[5] is cell A5. Similarly, u[1] is cell C1, u[2] is cell C2, u[5] is cell G1.

Then we can regard `t[i] = u[i] * 2` as shorthand for the equations

```
t[1] = u[1] * 2
...
t[5] = u[5] * 2
```

We can then interpret the indices in each equation as in the previous paragraph. This gives us the formulae:

```
A1 = C1 * 2
...
A5 = G1 * 2
```

This is how Excelsior works. Had I said that t was cells P77:P81, the process would have worked just as well; and it would have worked just as well if t and u held 9 or 127 elements instead of 5.

This gives us the "way to encode this scheme in a way independent of tables' sizes, positions and orientations" that Section 2.2 asked for. Components are encoded as Excelsior programs. Generating a group of cells from a component just entails inserting the user-wanted tables' sizes, positions and orientations into the program, running Excelsior over it, and sending the resulting spreadsheet to the user to copy into their own spreadsheet.

## 3. IMPLEMENTATION

Because the repository is a Web site, we had to design some Web pages as an interface. These are explained next, in Section 3.1. We also had to write code on the Web server to accept inputs from the component-customisation page, run Excelsior, and send back a link to the resulting downloadable file. This is explained in Section 3.2. These sections deal with the version for Excel. Section 3.3 briefly describes how the version for Google differs.

### 3.1 The Web pages

We designed four pages: a welcome page; the repository contents page; a page via which the user can request a component; and a download page.

### 3.1.1 The introduction page

This page welcomes the user, and explains what the repository can do for them. The underlined "contents list" in the second paragraph is a link:

> **Barking Dog Spreadsheet Parts Repository: Welcome**
>
> You are a spreadsheet developer. You are the Harry Potter of modelling: you know precisely how to model your company's business into Excel. But many Excel tasks require specialised knowledge. Some, such as statistical calculations, go badly wrong if you program them by a method that looks right but doesn't consider things such as numerical accuracy, rounding errors, and division by zero. Others, such as creating lists of dropdown menus options from bigger lists, are just plain hard.
>
> **We provide a Web-based library of parts to carry out such tasks!** Just look up your needs in our contents list, and choose which part you want. Then, using a Web form, tell our server where in your spreadsheet the part should take its inputs from, and where it should put its results. WE DO THE REST! Our server will AUTOMATICALLY reshape the part to fit your spreadsheet. It will then send it as a link which you can copy into your spreadsheet using a small VBA (Visual Basic for Applications) program.
>
> Should you be unsure about how a part works or how to use it, full documentation is available. All parts come with an example spreadsheet demonstrating usage, which you can download and try. And all parts come with fully cross-referenced documentation pages, which like the example spreadsheets, are linked from a part's contents entry.
>
> ***Dougal the barking dog says WHY KEEP A DOG AND BARK YOURSELF!?!***

### 3.1.2 The contents page

This is the catalogue of components, linked from the welcome page. In our prototype, there is one component, described after the single bullet point. A real-life repository would have many more, and would provide tools for searching the catalogue. The component is the one I explained in Section 2; the underlined "here"'s are links:

---

**Spreadsheet Parts Repository: Contents**

- Filter: removes all strings not matching a pattern.
  This part takes an input table of strings. It removes all strings that do not match a specified pattern, placing the others in the output table.
  Please click here to customise and download the part.
  Please click here for an example spreadsheet.
  Please click here to read the source code and explanatory commentary.

---

### 3.1.3 The form-submission page

This page contains a form where the user customizes the component, and is linked from the contents page above. Customisation involves telling the repository: where the component should take its inputs (the "elements to search" table of Section 2.1) from; where it should place its outputs (the "matching elements" table of Section 2.1); where the working table, "the index" in Section 2.1, should go; and the pattern to search for.

---

**Spreadsheet Parts Repository: Filter, Remove Non-matches**

This part takes an input table of strings. It removes all strings that do not match a specified pattern, placing the others in the output table. To get a copy of the part customised to your own spreadsheet, please fill in the fields below and click the Submit button.

Pattern to match

[            ]

| Input Sheet: | First input cell (top left) | Final input cell (bottom right) |
|---|---|---|
| [        ] | [        ] | [        ] |

| Output Sheet: | First output cell (top left) | Final output cell (bottom right) |
|---|---|---|
| [        ] | [        ] | [        ] |

Working Sheet:

[            ]

[Submit]

---

### 3.1.4 The "download customised component" page

The repository generates this page dynamically when the component is ready. It just contains a link to a temporary file generated by the server:

---

**Spreadsheet Parts Repository: Download**

Your component is ready. You can download it from <u>here</u>.

---

### 3.2 The server code

I implemented the system on my Internet Service Provider's (Mythic Beasts) machine, a fairly standard Unix box. Excelsior is written in SWI-Prolog [SWI], and fortunately, there is an SWI implementation for that machine and operating system. It can easily be called from Web scripting languages such as PHP, Perl and Python. Given these facts, implementation was straightfoward, using techniques familiar to any Web implementor. It worked as follows:

**1.** The user fills in the fields on the form described in the previous section and submits it. The Web browser then sends the fields' values to the Web server.

**2.** A script on the server, written in PHP, extracts the fields' values. From these, by some simple string manipulation, it generates a scratch file containing Excelsior statements defining the component's user-specifiable constants and cell locations. For example, the pattern to match would become a constant declaration such as `constant pattern = "X*"`. It appends this to the main Excelsior source file for the component. For the "Filter, Remove Non-matches" component, this is the file shown in Appendix 1. It then invokes Excelsior to compile this file.

**3.** By following the process described in Section 2.3, Excelsior generates a scratch text file containing the corresponding Excel formulae.

**4.** The server script emits a download page containing a link to this file.

**5.** The user downloads the file, and invokes a VBA macro to read the formulae from it into the appropriate cells of their spreadsheet.

### 3.3 The Google version

This works similarly. The key difference is that the user nominates the Google Spreadsheet to be updated, by filling in their Google Docs details and the spreadsheet name in a form [Paine 2008b]. They must also tell the repository where the component should be placed, as in Section 3.1.3. The repository then generates the formulae as in steps 2 and 3 above, but instead of putting them into a downloadable file, it copies them directly to the Google Spreadsheet. It does so by calling Google's Spreadsheets Data API (GData) library. At the lowest level, one has to manipulate Google Spreadsheets via HTTP requests [Google a], encoding information in URLs and as chunks of XML. The GData library hides the details of these encodings, and can be used without needing to know much about them.

## 4. EVALUATION

Evaluation was just a matter of testing that the implementation worked as expected. It did. Implementation often throws up unpleasant conceptual or technical surprises, but this time there were none. I did not evaluate the interface for ease of use; that should be one line of future research, using a more realistic catalogue of components.

## 5. FURTHER WORK

The most important task is to survey users to discover which components would be most useful. Then to implement them and promote the use of tested, documented, guaranteed, free components.

We should consider the user interface. With my prototype, users must re-request a component every time they change the size, shape, or location of its inputs or outputs in their spreadsheet. Can we design a user interface that handles this, whilst remaining acceptably unintrusive? (Popping up a "Do you want to reload component?" modal window every time the user edits any cell is not acceptably unintrusive.)

One possibility is that the user define named ranges for each input and output. An acolyte process could watch these ranges, automatically regenerating the component whenever they change.

Going further, we could program a "wiring tool" with which users could draw "wiring diagrams" depicting their spreadsheets as networks of interconnected components. Although such a tool would be primarily for use with components from the repository, it would also be usable with users' own components. (I shall say more in a future paper about how these will be created.) The "wiring diagrams" could then sit besides Excel as an alternative and more structured representation of the spreadsheet. Thus by stealth we encourage users to structure their spreadsheets.

We could consider using one of the Web-based spreadsheets for demonstrating example uses of components and perhaps for delivering to Excel. This was one reason I decided to experiment with a Google Spreadsheets version. It will be, of course, useful to Google Spreadsheets users anyway. Indeed, it would sit very nicely besides Google's own repository of charts and other data visualisation aids for spreadsheets, Google Gadgets [Chitu 2008].

## 6. CONCLUSIONS

I have prototyped a Web-based "spreadsheet component repository" from which users can copy "components" into their own Excel or Google spreadsheets. These components are collections of cells containing formulae that (in real life) would do calculations useful but too hard for many users to program. Crucially, the user can tell the repository which cells in their spreadsheet to use for a component's inputs and outputs. Because the repository stores components as "templates" (code for the Excelsior spreadsheet-generator) from which spreadsheets can be generated by parameterization, it can "reshape" the component to fit before sending it. A single component can therefore be used in many different sizes and shapes of spreadsheet.

I believe there to be two significant contributions to spreadsheet practice. Firstly, I hope to set up a spreadsheet equivalent of the high-quality numerical subroutine libraries that revolutionised scientific computing, but where instead of subroutines, the library contains "templates" for pieces of spreadsheet. This will bring to spreadsheeting the benefits that prefabrication of standardised parts brought to manufacturing.

Secondly, it seems that we can design an interface that will display spreadsheets as "wiring diagrams" for systems of interconnected components. Users may start off using this only for components downloaded from a repository. But once they are used to the idea, they may go on define and "wire in" their own components, structuring their own spreadsheets as collections of their own components. Thus, by stealth we encourage structured spreadsheets.

## APPENDIX 1: EXCELSIOR SOURCE AND DOCUMENTATION FOR "Filter, Remove Non-matches" COMPONENT

In the code below, the input cells are table `elements_to_search`, to be searched for occurrences of `pattern`.

The output cells are `matching_elements`. `matching_elements[i]` is the i'th element of `elements_to_search` that matches `pattern`, or blank if there are no more such elements.

The working cells are `the_index`. Element `the_index[i]` is the index of the i'th element of `elements_to_search` that matches `pattern`, or -1 if there are no more such elements.

```
  constant pattern.

  table elements_to_search : elements_base -> text.

  table matching_elements : elements_base -> text.

  type elements_base.

  table the_index : elements_base -> text.

  the_index[ 1 ] =
    if( isna( match( pattern, elements_to_search[all], 0 ) )
      , -1
      , match( pattern, elements_to_search[all], 0 )
      ).

  the_index[ i > 1 ] =
    if( the_index[i-1] = -1
        // There was no previous one, so can't be another.
      , -1
      , if( the_index[i-1] = upb(elements_base)
            // Previous one was at the end of the table,
            // so there can't be any more.
          , -1
          , if( isna( match( pattern
                           , elements_to_search[ (the_index[i-1]+1):upb(elements_base) ]
                           , 0
                           )
                    )
              , -1
              , match( pattern
                     , elements_to_search[ (the_index[i-1]+1):upb(elements_base) ]
                     , 0
                     ) + the_index[ i-1 ]
              )
          )
      ).

  matching_elements[ i ] =
    if( the_index[ i ] <> -1
      , elements_to_search[ the_index[i] ]
      , ""
      ).
```

## APPENDIX 2: EXCELSIOR SOURCE AND DOCUMENTATION FOR SF STORY SPREADSHEET

This Appendix is a longer example, demonstrating specification and documentation of an unusual Excel spreadsheet, one that generates science-fiction plots. The point is to show Excelsior-style documentation at work on a task that almost all readers will find unfamiliar. If the documentation makes it clear to you, it will help to show Excelsior to be a viable documentation tool.

The text below is how Excelsior's documentation engine renders the text as HTML: for details, see [Paine 2007a]. For the print copy of this paper, I have removed the links between tables since they don't show up on paper. I have also indented the code sections to make them stand out. On the Web, they would stand out anyway, because the documentation engine places them on a pale blue background. The actual HTML generated, and the spreadsheet, can be seen at [Paine 2007b].

This is an Excelsior version of the Prolog science-fiction generator at www.j-paine.org/cgi-bin/spin.php. I adapted that somewhat broadly from [Wilson 1972]. It demonstrates how recursion is possible when one thinks of tables as functions.

### Defining the content of possible stories

The raw material from which I generate stories is a graph — that is, a network — of possible story events. The graph consists of nodes (i.e. points), and edges, which lead from one node to another.

Each edge is associated with a chunk of text. A node can have more than one edge leading from it. To generate a story, we start at node 0 and select, at random, one of the three edges leading from it. The text associated with the selected edge becomes the first few words of the story.

Every edge leads to a node; and so the edge selected from node 0 will too. To generate the rest of the story, we repeat the above process with that node; and we simply continue following edges and using their associated text until we arrive at a node from which no edges lead.

### Representing the graphs in Excel

#### Nodes: table `node_nos`

I represent each node by an integer, starting at 0. The data stored in this version of the spreadsheet has nodes numbered from 0 to 42:

```
type node_base = 0:42.
```
I'm going to set up a table that lists the edges leading out of each node. To make it easier to read, I'll put it next to a table that shows all the node numbers — i.e. the `n`'th element of the table is just `n`:

```
  table node_nos : node_base -> general.

  node_nos[n] = n.
```

#### Edges: table `out_edges`

Now I'll define the edges that lead out of each node. For each edge, I need to store its text, and the number of the node it leads to. I must do so in a way that makes it easy to count the edges, and to randomly select one of them, and to find the node it leads to.

It must also be easy to type new text and node numbers when defining the story data.

Because Excel's string-handling is rudimentary, I decided to store the text of an edge and the number of the node it leads to in separate but adjacent cells, rather than in one cell which would have to be chopped apart to get these components. So each edge is two cells, text followed by a destination number. I'll store all the edges for a node in the same row, so that a row becomes a sequence of `Text DestinationNodeNumber` cell pairs. This way, if I generate a random number to indicate an edge, it's easy to get the corresponding text and destination node number.

So I want a two-dimensional table, whose first dimension is node number, and whose second dimension is edge data:

```
  type out_edges_base = 0:12.

  table out_edges : node_base out_edges_base -> general.
```

This will contain the story data. I've put the data at the end of this file; to show you how it works, here are the first few lines:

```
  // out_edges[0,0]="Earth".
  // out_edges[0,1]=1.
  // out_edges[0,2]="Mars".
  // out_edges[0,3]=1.
  // out_edges[0,4]="Planet 9 of Alpha-Centauri".
  // out_edges[0,5]=1.
  // out_edges[1,0]="is used as the cue ball in a game of galactic
bar-billiards".
  // out_edges[1,1]=2.
  // out_edges[1,2]="falls toward the Sun".
  // out_edges[1,3]=2.
  // out_edges[1,4]="falls toward a black hole".
  // out_edges[1,5]=2.
```

**Counting the edges: table `out_edge_count`**

As already mentioned, to generate a story, we pick the start node, randomly select an edge leading from it, output the associated text, and repeat with the edge's destination node.

To do the random selection, I use the Excel function `rand()`, asking it to generate a number between 0 and the number of edges minus 1. It turned out to be convenient to simplify the formula by calculating this number of edges in an auxiliary table:

```
  table out_edge_count : node_base -> general.

  out_edge_count[ n ] = count( out_edges[n, 0:12] ).
```

**The generated story**

I'll explain what I did as though the story is being generated over a sequence of timepoints, with the first node generated at time 0. Let's have 100 timepoints altogether:

```
  type story_base = 0:99.
```
The core of this representation will be the node number selected at each timepoint.

**Story nodes: table `story_node_nos`**

```
  table story_node_nos : story_base -> general.
```

We start with node number 0, at time 0:
```
story_node_nos[ 0 ] = if( go[1] = 'Recalculate', 0, 0 ).
```
(The reference to table `go` enables the user to force generation of a new story by selecting from a dropdown at the top of the spreadsheet.)

For the remaining timepoints, the story node at time `t` will be the node to which the edge selected at time `t-1` leads:

```
story_node_nos[ t>0 ] = story_out_edge_destination_node_no[ t-1 ].
```

**Story destination nodes: `story_out_edge_destination_no`**

The right-hand side of that equation introduces a new table which holds the number of the node selected at each time:

```
table story_out_edge_destination_node_no : story_base -> general.
```
I'll define this in terms of an auxiliary index. Recall that I've stored a node's out-edges in a row, as pairs of cells: Text DestinationNodeNumber Text DestinationNodeNumber Text DestinationNodeNumber ...

Suppose we have selected pair number `i`: `i` is 0 for the first pair, and so on. Then the `i`'th pair's text is offset by `i*2` from the first cell in this row. The `i`'th pair's destination node number is offset by `i*2 + 1`.

So now suppose that we have a table `story_out_edge_index`, giving the value of this `i` at time `t`. Then we can say that:

```
// story_out_edge_destination_node_no[ t ] =
// offset( out_edges[0,0], story_node_nos[ t ],
story_out_edge_index[ t ]*2+1 ).
```

**When we crash into the end of the story**

Note that I've commented out the above equation. The one I'm actually using is this:

```
story_out_edge_destination_node_no[ t ] =
  if( story_out_edge_index[ t ] <> -1
    , offset( out_edges[0,0], story_node_nos[ t ],
story_out_edge_index[ t ]*2+1 )
    , -1
    ).
```
It's as above, but returns -1 if the index is -1. I'm using this to handle the case where I've run out of story nodes — i.e. after coming to a node with no out-edges. It's important to deal with such cases, otherwise we'll end up with cellsful of #REF's.

**The index: table `story_out_edge_index`**

This is where the random edge selection happens. The index of the out-edge selected at time `t` is -1 if there are no edges, otherwise it's a random number between 0 and the number of edges minus 1:

```
table story_out_edge_index : story_base -> general.
```

```
story_out_edge_index[ t ] =
  if( story_out_edge_count[ t ] > 0
    , floor( rand() * story_out_edge_count[ t ], 1 )
    , -1
    ).
```

Note that 'rand' is better than 'randbetween', because the latter requires the Excel Analysis ToolPak to be loaded. Thanks to Alice Campbell for pointing out this improvement.

**The out-edge count: table `story_out_edge_count`**

I used an auxiliary table to get the number of out-edges at time `t`. It's just the number of out-edges for the corresponding node, or -1 if we've run out of story so the node is -1:

```
table story_out_edge_count : story_base -> general.

story_out_edge_count[ t ] =
  if( story_node_nos[ t ] <> -1
    , offset( out_edge_count[0], story_node_nos[ t ], 0 )
    , 0
    ).
```

**The text associated with an out-edge: `story_out_text`**

This uses the index of the selected out-edge, as explained earlier, to get the edge's text:

```
table story_out_text : story_base -> general.

story_out_text[ t ] =
  if( story_out_edge_index[ t ] <> -1
    , offset( out_edges[0,0], story_node_nos[ t ], story_out_edge_index[ t ]*2 )
    , -1
    ).
```

**The output: table `story_node_text`**

This appears near the top of the spreadsheet and is held in the table `story_node_text`:

```
table story_node_text : story_base -> text.
```
It's a copy of `story_out_text`, cleaned up to display blanks after we've hit the end of the story:
```
story_node_text[ t ] =
  if( story_out_text[ t ] <> -1
    , story_out_text[ t ]
    , ""
    ) &

  if( and( t = upb( story_base )
        , story_out_edge_count[ t ] > 0
        )
    , " ... NO SPACE TO CONTINUE!"
    , ""
    ).
```
The second `if` appends a suitable message if we're about to run off the end of the table.

**Forcing recalculation**

As explained above, this forces recalculation. In the layout, I've attached a dropdown to this table.

```
table go : -> general.
```

**Layout**

```
  layout( 'Spin'
       , rows( [ skip(1,2) ]
             , [ 'Story', 'Recalculate' ]
             , [ story_node_text as y, as( go, y, ['Recalculate']
) ]
             , [ skip(1,2) ]
             , heading
             , [ story_node_nos as y
               , story_out_edge_count as y
               , story_out_edge_index as y
               , story_out_edge_destination_node_no as y
               , story_out_text as y
               ]
             , [ skip(1,2) ]
             , heading
             , [ node_nos as y
               , out_edges as yx
               , out_edge_count as y
               ]
             )
       ).
```

**Story data**

```
  out_edges[0,0]="Earth".
  out_edges[0,1]=1.
  out_edges[0,2]="Mars".
  out_edges[0,3]=1.
  out_edges[0,4]="Planet 9 of Alpha-Centauri".
  out_edges[0,5]=1.

  out_edges[1,0]="is used as the cue ball in a game of galactic
bar-billiards".
  out_edges[1,1]=2.
  out_edges[1,2]="falls toward the Sun".
  out_edges[1,3]=2.
  out_edges[1,4]="falls toward a black hole".
  out_edges[1,5]=2.
  out_edges[1,6]="is struck by a comet".
  out_edges[1,7]=2.
  out_edges[1,8]="is invaded by nasty aliens".

  ... etc.

  out_edges[28,0]="".
  out_edges[28,1]=-1.
  out_edges[28,2]="and".
  out_edges[28,3]=22.
  out_edges[42,0]="42".
```

```
out_edges[42,1]=13.
```

**APPENDIX 3: THE EXCELSIOR LANGUAGE**

This Appendix explains just enough of the Excelsior language for you to make sense of Appendices 1 and 2.

Excelsior describes spreadsheets as sets of equations between groups of cells called tables. The Excelsior compiler reads program files containing the equations and generates an Excel spreadsheet by translating each equation into one or more formulae.

As well as equations, a program can contain constant declarations, type declarations, table declarations, and layout descriptions. The program's meaning is independent of their order. Programs are free format; statements end with a full stop; and Excelsior recognises /* to */ as block comments and // as end-of-line comments.

Each equation has a left-hand side and a right-hand side. The right-hand side is an Excel formula, except that instead of cell references, it contains Excelsior table references. It can also refer to declared constants. The left-hand side is a reference to a table element. To translate the equation, Excelsior rewrites the right-hand side, replacing table references by cell references, and constants by their values. It then places the formula into the cell denoted by the left-hand side.

Excelsior calculates cell positions from layout statements, each of which depicts a sheet as a grid of tables. In the example below, there is one layout statement, depicting sheet Demo.

The following contrived program illustrates this:

```
  constant two = 2.
          // An uninteresting constant.

  constant hundred = 100.
          /* A bigger constant. */

  type span = 1:4.

  table nums: span -> general.
          // I'll put numbers
          // in here. There'll be 4, because
          // 'span' is 1:4.

  table strings: span -> text.
          /* I'll put messages in here. */

  nums[1] = two.
  strings[1] = "Two = " & two & ".".

  nums[2] = two * nums[1].
  strings[2] = "Twice two = " & nums[2] & ".".

  nums[3] = len( strings[2] ).
  strings[3] = "Length of above text = " & nums[3] & ".".

  nums[4] = sum( nums[1], nums[2:3], hundred, 250.12+249.88 ).
```

```
strings[4] = "Sum of above numbers plus 600 = " & nums[4] & ".".

layout( 'Demo'
      , rows( [ nums, strings ] )
      ).
         // This makes sheet 'Demo' contain
         // 'nums' to the left of 'strings'.
```

This program generates one sheet, called `Demo`. This contains tables `nums` and `strings`. Both tables contain 4 elements, as specified by the type span used in their declarations. When arranged on sheet `Demo`, `nums` is in column A and to the left of `strings`, as specified in the layout statement. The spreadsheet generated from the above is as follows:

| 2   | Two = 2.                              |
|-----|---------------------------------------|
| 4   | Twice two = 4.                        |
| 14  | Length of above text = 14.            |
| 620 | Sum of above numbers plus 100 = 620.  |

The table declarations, incidentally, have a syntax copied from the convention often used to describe functions in mathematics and computer science: the function's name, followed by the types of its arguments, followed by the type of its result. This is because it's often convenient to think of tables as functions, and table references as function calls. The story-generator in Appendix 2 illustrates this: by recursing over tables, it finds a random path through a transition matrix represented as a graph.

In Excelsior, the argument types specify the size of a table's dimensions. The result type specifies the type of the elements the table should contain. Excelsior uses this both to detect errors such as adding strings to numbers, and to generate an appropriate Excel style for formatting the cells.